\documentclass[%
 reprint,
superscriptaddress,
groupedaddress,
showpacs,
preprintnumbers,
nofootinbib,
nobibnotes,
 amsmath,
 amssymb, 
 aps,
 prl,
floatfix
]{revtex4-1}

\usepackage[utf8]{inputenc}
\usepackage[dvipsnames]{xcolor}
\usepackage[normalem]{ulem}
\usepackage{graphicx}% Include figure files
\usepackage{dcolumn}% Align table columns on decimal point
\usepackage{bm}% bold math
\usepackage[colorlinks=true,allcolors=magenta]{hyperref}
\usepackage{url}

\usepackage{slashed,multirow,relsize,soul,feynmp-auto,tikz}
\usepackage{color}
\usepackage{mathrsfs} % pretty maths
\newcommand{\refeq}[1]{Eq.~(\ref{#1})}

\newcommand{\reffig}[1]{Fig.~\ref{#1}}

\def\eg{\emph{e.g.}}
\def\ie{\emph{i.e.}}

\newcounter{CommentCount}
\definecolor{PB}{rgb}{0.9,0,0}
\definecolor{MH}{rgb}{0.0,0.9,0}
\definecolor{SP}{rgb}{0.0,0.0,0.9}

\definecolor{palatinate}{rgb}{0.494, 0.192, 0.482}

\begin{document}

\preprint{\hfill IPPP/19/21}

% \title{Neutrino Masses from a Light Dark Neutrino Sector}% Force line breaks with \\
\title{Neutrino Masses from a Dark Neutrino Sector below the Electroweak Scale}% Force line breaks with \\

\author{Peter Ballett}
\author{Matheus Hostert}
\email{matheus.hostert@durham.ac.uk}
\author{Silvia Pascoli}
\email{silvia.pascoli@durham.ac.uk}

\affiliation{Institute for Particle Physics Phenomenology, Department of
Physics, Durham University, South Road, Durham DH1 3LE, United Kingdom.}

\date{\today}% It is always \today, today,

\begin{abstract}

We consider a minimal extension of the Standard Model which advocates a dark neutrino sector charged under a hidden $U(1)^\prime$. We show that neutrino masses can arise radiatively in this model. The observed values are compatible with a light dark sector below the electroweak scale and would imply new heavy fermions which may be testable in the next generation of beam dump searches at DUNE, NA62 and SHIP.

\end{abstract}

\maketitle

%%%%%%%%%%%%%%%%%%%%%%%%%%%%%%%%%%%%%%%%%%%%%%%%%%%%%%%%%%%%%%%%%%%%%%%
\section{Introduction} Neutrino oscillations have been established by several experiments~\cite{Fukuda:1998ah,*Ahmad:2002jz,*Eguchi:2002dm}, implying small but non-vanishing neutrino masses. In the Standard Model (SM), neutrinos are strictly massless due to the absence of right-handed neutrino fields, urging for extensions of the theory. The Type-I seesaw mechanism~\cite{Minkowski:1977sc,*Mohapatra:1979ia,*GellMann:1980vs,*Yanagida:1979as,*Lazarides:1980nt,*Mohapatra:1980yp,*Schechter:1980gr,*Cheng:1980qt,*Foot:1988aq}, arguably the most popular mechanism to explain the lightness of neutrino masses, relies on the addition of at least 2 heavy right-handed neutrinos $N_R$. The large scales of $N_R$ and/or the smallness of the Yukawa couplings makes the minimal realisation of this model difficult to test. Therefore, searching for variations of the Type-I seesaw where novel and testable phenomena are present is an essential part of solving the neutrino mass puzzle~\cite{Boucenna:2014zba}. A few notable examples of such alternatives are the Inverse Seesaw (ISS)~\cite{Mohapatra:1986bd,*GonzalezGarcia:1988rw} and the Linear Seesaw (LSS)~\cite{Wyler:1982dd,*Akhmedov:1995ip,*Akhmedov:1995vm}, where the lightness of neutrino masses is explained by an approximate conservation of lepton number, and the Extended Seesaw (ESS)~\cite{Barry:2011wb,*Zhang:2011vh}, where new heavy neutral fermions generally appear at small scales.
This class of models assumes additional SM gauge neutral fermions that mix with light neutrinos, usually referred to as sterile neutrinos. These, however, need not be completely sterile and might have new gauge interactions shared with the SM fermions~\cite{%
Buchmuller:1991ce,% original
Khalil:2006yi,% B-L at TeV scale
Perez:2009mu,% B-L extensive discussion + X model
Khalil:2010iu,% B-L at collider and ISS
Dib:2014fua,% B-L in linear seesaw
Baek:2015mna,% mu-tau in inverse seesaw
DeRomeri:2017oxa,% Elusive B-L
Nomura:2018mwr,% U(1)_R
Brdar:2018sbk% LR symmetry low scale
} or not~\cite{%
Okada:2014nsa,% U(1)prime
Diaz:2017edh,% Hidden witloops
Bertuzzo:2017sbj,% nu2HDM
Nomura:2018ibs,% Hidden in linear seesaw
Bertuzzo:2018ftf% dark neutrinos PM
}. In the latter case, we refer to these new heavy fermions as \textit{dark neutrinos}. The interest in such particles arises from their novel interactions which may ``leak" into the SM sector via neutrino mixing, where they offer a variety of phenomenological and cosmological consequences. 

In this article, we consider the new minimal model introduced in Ref.~\cite{Ballett:2019pyw}.
It introduces two type of new neutral fermions, namely dark neutrinos $\nu_D$ and additional sterile neutrinos $N$. We impose a hidden $U(1)^\prime$ gauge symmetry with the associated hidden gauge boson $X_\mu$, which mediates the dark neutrino interactions. The symmetry is subsequently broken by the vacuum expectation value (vev) of a complex dark scalar $\Phi$. As discussed in Ref.~\cite{Ballett:2019pyw}, the model can exhibit a significantly different phenomenology than the case of neutrino mixing only. Beyond evading many current bounds, such dark neutrinos could explain the MiniBooNE anomaly as discussed in~\cite{Ballett:2018ynz} (see also~\cite{Bertuzzo:2018itn}) and lead to novel neutrino scattering signatures~\cite{Arguelles:2018mtc}. Bounds on dark photons might also be severely weakened. If kinematically allowed, they would mainly decay into heavy neutrinos, which may be invisible or lead to multi-lepton plus missing energy signatures. 

In this article, we discuss the generation of neutrino masses in our dark neutrino model.
Crucially, the new gauge symmetry forbids Majorana mass terms for the $\nu_D$ states and, after symmetry breaking, leads to a mass matrix similar to the one in the so-called Minimal ISS~\cite{Dev:2012sg}. As such, this symmetry-enhanced seesaw predicts vanishing light neutrino masses at tree-level. Here, we show that it induces their radiative generation via one-loop diagrams involving the new scalar and vector particles~\cite{Dev:2012sg,Zhang:2013ama,Diaz:2017edh}. After identifying the range of heavy neutrino parameters required to explain the observed light neutrino masses, we point out interesting phenomenological consequences.

% DIAGRAMS
%%%%%%%%%%%%%%%%%%%%%%%%%%%%%%%%%%%%%%%%%%%%%%%%%%%%%%%%%%%%%%%
\begin{figure*}[t]
\centering
\includegraphics[width=\textwidth]{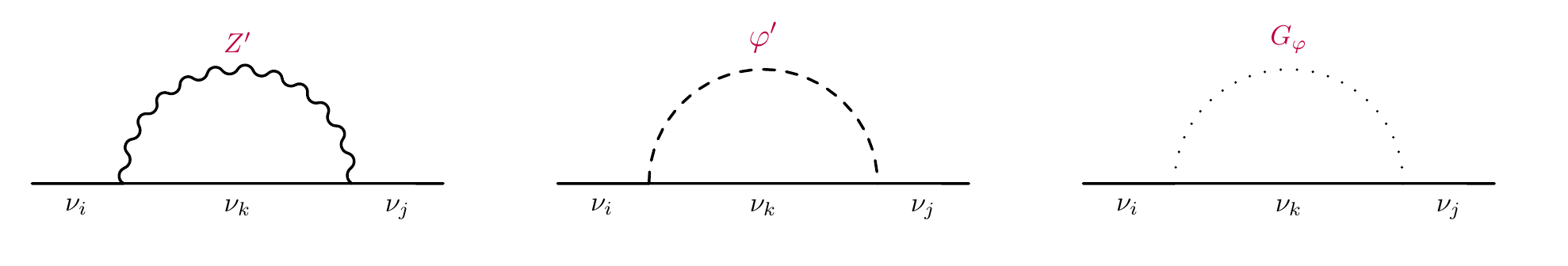}
\caption{\label{fig:loops}The three contributions to the neutrino self-energy arising from novel bosons in the theory.}
\end{figure*}
%%%%%%%%%%%%%%%%%%%%%%%%%%%%%%%%%%%%%%%%%%%%%%%%%%%%%%%%%%%%%%%

\section{Model set-up} Following~\cite{Ballett:2019pyw}, we add two types of heavy neutral fermions to the SM, namely a dark neutrino $\nu_{D,L} \equiv \nu_{D}$ and a sterile state $N_L \equiv N$. For simplicity, we restrict the discussion to one generation in order to focus on the main features of the model.

We impose a new abelian gauge symmetry $U(1)^\prime$ with associated mediator $X_\mu$ and introduce a neutral complex scalar $\Phi$. No SM fields are charged under $U(1)^\prime$. The scalar $\Phi$ and the fermion $\nu_{D}$ carry the same $U(1)^\prime$ charge, while $N$ remains completely neutral. The gauge-invariant Lagrangian is given by
\begin{align} \label{eq:lagrangian}
\mathscr{L} =& \mathscr{L}_{\mathrm{SM}} + \left(D_\mu \Phi\right)^\dagger \left(D^\mu \Phi\right) -  V(\Phi,H) \,   \nonumber\\
&  - \frac{1}{4}X^{\mu \nu} X_{\mu \nu} + \overline{N}i\slashed{\partial}N + \overline{\nu_D}i\slashed{D}\nu_D 
 \nonumber\\
&- \left[y^\alpha_\nu (\overline{L_\alpha} \cdot \widetilde{H})N^c + \frac{\mu^\prime}{2}\overline{N}N^c + y_N \overline{N}\nu_D^c\Phi + \text{h.c.}\right],
\end{align}
where $X^{\mu\nu} \equiv \partial^\mu X^\nu - \partial^\nu X^\nu$, $D_\mu \equiv \left(\partial_\mu-ig^\prime X_\mu\right)$, $L_\alpha \equiv (\nu_\alpha^T, \ell_\alpha^T)^T$ is the SM leptonic doublet of flavour $\alpha = e, \mu, \tau$ and $\widetilde{H} \equiv i \sigma_2 H^*$ is the charge conjugate of the SM Higgs doublet. In the neutral fermion sector, we have Yukawa couplings $y_\nu^\alpha$ and $y_N$  responsible for $L_\alpha$-$N$ and $\nu_D$-$N$ interactions, respectively, and a Majorana mass $\mu^\prime$ for $N$. The latter term violates by two units any lepton number assignment which leaves 
the Yukawa term $L_\alpha$-$N$ invariant. As such, it plays a crucial role in the generation of light neutrino masses, as we discuss.

We are interested in the case in which both the neutral component of the fields $H$ and $\Phi$ acquire non-vanishing vevs, $v_H$ and $v_\varphi$. They induce mixing between active and heavy fermions, and give a mass to the gauge boson $X_\mu$ and to the real component of the scalar field $\varphi$. We are interested in proposing a model for neutrino masses which is testable in current and future non-collider experiments, and as such we focus on a new physics scale which is below the electroweak one, $v_\varphi < v_H$.  
In addition to the neutrino portal, this model can accommodate a vector portal arising from vector kinetic mixing term and a scalar portal coming from the cross-coupling term $H^\dagger H \Phi^\dagger \Phi$ in the potential~\cite{Ballett:2019pyw}. Kinetic mixing can be reabsorbed in a redefinition of vector fields, leading to a new gauge boson which has vector couplings to the SM fermions proportional to their electric charge. For our neutrino mass generation mechanism, the vector and scalar mixing do not play a relevant role and we set them to zero from here onward, unless otherwise specified. Regarding the vector boson, we refer to it as a $Z^\prime$, independently of kinetic mixing.

%%%%%%%%%%%%%%%%%%%%%%%%%%%%%%%%%%%%%%%%%%%%%%%%%%%%%%%%%%%%%%%%%%%%%%%
\section{Neutrino masses}  
After symmetry breaking, two Dirac mass terms are induced with $m_D \equiv y_\nu^\alpha v_H/\sqrt{2}$ and $\Lambda \equiv y_N v_\varphi/\sqrt{2}$.
For one active neutrino $\nu_\alpha$, $\alpha= e, \mu, \tau$, the mass matrix is given by
\begin{align} \label{eq:massmatrix}
\mathscr{L}_{\rm mass} \supset
\frac{1}{2}\left (\begin{matrix} \overline{\nu}_\alpha & \overline{N} &  \overline{\nu_D} \end{matrix} \right )
\left(\begin{matrix} 
     0   &  m_D        & 0 
\\ m_D &  \mu^\prime & \Lambda 
\\   0   &  \Lambda  & 0
\end{matrix}\right)
\left (\begin{matrix} \nu_\alpha^c \\ N^c \\ \nu_D^c \end{matrix} \right) + {\rm h.c.}
\end{align}  
Let us emphasize the fact that in our model the zeros in the $\nu_D$-$\nu_D$ and $\nu_\alpha$-$\nu_D$ entries are enforced by the $U(1)^\prime$ symmetry, differently from LSS and ISS models, in which these are generically assumed to be nonzero and small due to the quasi-preservation of lepton number. Here, lepton number violation (LNV) may be large, as the $\mu^\prime$ term breaks it by 2 units. Alternatively, it can be small and technically natural, leading to quasi-degenerate heavy neutrinos, see below. 

After diagonalisation of the mass matrix, the two heavy neutrinos,  $\nu_h$ ($h=4,5$), acquire masses
\[m_{4,5} = \frac{\mu^\prime \mp \sqrt{\mu^{\prime\,2} + 4 (\Lambda^2 + m_D^2) } }{2}.\]
Assuming that $m_D \ll \Lambda$, we focus on two interesting limiting cases. 

The ISS-like scenario is defined by $\Lambda \gg \mu^\prime$: the two heavy neutrinos are nearly degenerate with a mass $\Lambda$ and mass splitting $\mu^\prime$. The relevant mixing parameters are $U_{\alpha 4,5} \sim m_D/\sqrt{2} \Lambda$ and $U_{D4,5} \sim 1/\sqrt{2}$.
The ESS-like case has $\Lambda \ll \mu^\prime$: one neutral lepton remains very heavy, $m_5 \sim \mu^\prime$, and mainly in the completely neutral direction $N$, and the other acquires a small mass via the seesaw mechanism in the hidden sector with $m_4 \sim - \Lambda^2/\mu^\prime$ and $U_{D5} \sim \Lambda/ \mu^\prime$. The mixing with active neutrinos is given by $U_{\alpha 5} \sim m_D/\mu^\prime \ll U_{\alpha 4} \sim m_D/\Lambda$.

The specific form of the mass matrix in Eq.~\ref{eq:massmatrix} implies vanishing light neutrino masses at tree level, as its determinant is zero~\cite{Dev:2012sg,LopezPavon:2012zg}. This feature holds to all orders in the seesaw expansion~\cite{Grimus:2000vj,Adhikari:2010yt,LopezPavon:2012zg}. The light neutrino masses, however, are not protected by any symmetry and arise from radiative corrections (for a review of radiative neutrino mass models see, \eg, Ref.~\cite{Cai:2017jrq}).

%%%%%%%%%%%%%%%%%%%%%%%%%%%%%%%%%%%%%%%%%%%%%%%%%%
\subsection{Radiative corrections} We now show that our model generically leads to the generation of light neutrino masses at one loop. The calculation of the radiative mass term follows Refs.~\cite{Pilaftsis:1991ug,Kniehl:1996bd} with the addition of the loops with the new boson and scalar particles shown in \reffig{fig:loops}. The self-energy of the Majorana neutrino fields is given by 
\begin{equation*}
 \Sigma_{ij}(\slashed{q}) = \slashed{q}P_\text{L}\Sigma^\text{L}_{ij}(\slashed{q}) + \slashed{q}P_\text{R}\Sigma^\text{L*}_{ij}(\slashed{q}) + P_\text{L}\Sigma^\text{M}_{ij}(q^2)+ P_\text{R}\Sigma^{\text{M}*}_{ij}(q^2).
\end{equation*}
Using the on-shell renormalization scheme, the renormalized mass matrix for the light neutrinos, massless at tree level, emerges at one-loop and is given by~\cite{Kniehl:1996bd}
\[    m_{ij}^\text{one-loop} = \text{Re}\left[ \Sigma^\text{M}_{ij}(0)\right], \quad  i, j <4. \]
The self energy can be decomposed as
\begin{align}
\Sigma_{ij}^\text{M}(0) = \Sigma^Z_{ij}(0)& + \Sigma^{h}_{ij}(0) + \Sigma^{G_h}_{ij}(0) \,+ \nonumber\\ &\Sigma^{Z^\prime}_{ij}(0) +
\Sigma^{\varphi^\prime}_{ij}(0) + \Sigma^{G_\varphi}_{ij}(0),
\end{align}
where $\Sigma^{Z, h, G_h}$ come from the SM particles, $Z^0$, the Higgs and the associated Goldstone boson, respectively, and $\Sigma^{Z^\prime, \varphi^\prime, G_\phi}$ are the new terms present in our model, mediated by the new gauge boson and new scalar components. From it, we write the $3\times3$ light neutrino mass matrix 
\begin{align}\label{eq:masses_general}
 m_{ij} = \frac{1}{4\pi^2}\sum_{k=4}^5 \Big[ & C_{ik} C_{jk} \frac{m_k^3}{m_Z^2}F(m_k^2,m_Z^2,m_h^2)  \,+ 
 \nonumber\\ &D_{ik} D_{jk} \frac{m_k^3}{m_{Z^\prime}^2}F(m_k^2,m_{Z^\prime}^2,m_{\varphi^\prime}^2) \Big], 
\end{align} 
where we defined coupling matrices corresponding to the SM and new physics interaction terms assuming $\chi=\lambda_{\Phi H}=0$:
\begin{equation} \label{eq:couplings}
C_{ik} \equiv \frac{g}{4c_W}\sum_{\alpha = e}^\tau U_{\alpha i}^*U_{\alpha k}\quad\text{and} \quad D_{ik} \equiv \frac{g^\prime}{2} U^*_{Di} U_{Dk}.
\end{equation}
Equivalent expressions can be found for non-vanishing portal couplings, but considering experimental constraints we find that these do not play a role in the neutrino mass generation. It is possible to show that in general $\sum_{k} m_k C_{ik} C_{jk} =0$ and $\sum_{k} m_k D_{ik} D_{jk} =0$ for any $i,j$. By virtue of the latter property, the loop function can be written as
\begin{equation} \label{eq:loop_function}
F(a,b,c) \equiv \frac{3 \, \ln{(a/b)}}{a/b - 1}  + \frac{\ln{(a/c)}}{a/c - 1}.
\end{equation}
Turning off the $g^\prime$ gauge coupling, we recover the expression for the Type-I seesaw case~\cite{Pilaftsis:1991ug}:
 \begin{align}  \label{eq:SM_masses}
 m_{ij} = &\frac{\alpha_W}{16\pi}\sum_{\alpha, \beta = e}^{\tau} U_{\alpha i}^\ast U_{\beta j}^\ast  U_{\alpha 5} U_{\beta 5} \frac{m_5}{m_W^2} \times\nonumber\\ & \left( m_5^2 F(m_5^2,m_Z^2,m_h^2) -  m_4^2 F(m_4^2,m_Z^2,m_h^2)\right).
 \end{align}
These SM corrections to neutrino masses also arise in the Minimal ISS model~\cite{Dev:2012sg,LopezPavon:2012zg}. In the latter, however, no explanation is provided as to why they dominate neutrino masses. Moreover, if we restrict the discussion to scales well below the electroweak one, $ m_5 \ll 10$~GeV, bounds on the mixing angles severely constrain the parameter space viable to generate the observed values of the masses. 

For a light $Z^\prime$, the second term in Eq.~\ref{eq:masses_general} dominates
\begin{align}\label{eq:BSM_masses}
m_{ij} \simeq  &\frac{g^{\prime2}}{16\pi^2} U_{D i}^{*} U_{D j}^{*} \,  U_{D5}^2 \frac{m_5}{m_{Z^\prime}^2} \times\nonumber\\  \quad \quad & \big(m_5^2 F(m_5^2,m_{Z^\prime}^2,m_{\varphi^\prime}^2) - m_4^2 F(m_4^2,m_{Z^\prime}^2,m_{\varphi^\prime}^2)\big)~.
\end{align}
We notice that the resulting mass matrix has only one nonzero eigenvalue. This suggests that a typical prediction of our model is a normal ordering mass spectrum, in which $m_3$ is given by this radiative mechanism and $m_2$ has another origin, for example the loops mediated by the SM gauge bosons or by additional particle content. Our simplifying assumption of one generation of hidden fermions is by no means necessary and more generations of new fermions are possible, leading to a much richer structure for the light neutrino mass matrix. The additional $\mu^\prime$ terms would not be constrained and could be at different scales, while the $\Lambda$ terms arise from the $U(1)^\prime$ breaking and are therefore constrained to be at/below $v_\varphi$. Therefore, the full model could present a combination of relatively light Majorana $\nu_h$, mainly in dark direction, some very heavy nearly-neutral neutrinos and pseudo-Dirac pairs at intermediate scales. A discussion of this extension is beyond our scope, but we note that it has interesting consequences for both the heavy and light neutrino mass spectra and mixing structure.

Working in a single family case, we derive expressions for Eq.~\ref{eq:BSM_masses} in the seesaw limit for both the ISS and ESS-like scenarios. %%%%%%%%%%%%%%%%%%%%%%%%%%%%%%%%%%%%%%%%%%%%%%%%%%%%
% ISS
In the ISS-like regime and assuming $m_{Z^\prime}, m_{\varphi^\prime} \ll \, \Lambda$, \refeq{eq:BSM_masses} simplifies to
\begin{align} \label{eq:ISSlimit_1}
m_3 \simeq& \frac{g^{\prime 2}}{8\pi^2} \frac{m_D^2}{m_{Z^\prime}^2} \mu^\prime  \left( {3 \ln{\frac{m_{Z^\prime}^2}{\Lambda^2}} + \ln{\frac{m_{\varphi^\prime}^2}{\Lambda^2}} - 4 }\right),
\end{align}
while for $m_{Z^\prime}, m_{\varphi^\prime} \gg \, \Lambda$ it reduces to 
\begin{align} \label{eq:ISSlimit_2}
m_3 \simeq& \frac{g^{\prime 2}}{16\pi^2} \frac{m_D^2}{\Lambda^2} \mu^\prime \left( 3 + \frac{m_{\varphi^\prime}^2}{m_{Z^\prime}^2} \right).
\end{align}
As it can be expected, neutrino masses are controlled by the LNV parameter $\mu^\prime$ and are enhanced with respect to the SM contribution by a factor of $(m_Z/m_{Z^\prime})^2$ in the former, or $(m_Z/\Lambda)^2$ in the latter case. 

For the ESS-like regime, taking $m_{Z^\prime}, m_{\varphi^\prime} \ll \mu^\prime$, the light neutrino mass is approximately
\begin{align} \label{eq:ESSlimit_1}
m_3 \simeq& \frac{g^{\prime \,2}}{16\pi^2} \frac{m_D^2}{\Lambda^2 + m_D^2} \frac{\Lambda^2}{m_{Z^\prime}^2} \mu^\prime  \left( {3 \ln{\frac{m_{Z^\prime}^2}{ \mu^{\prime 2} }} + \ln{\frac{m_{\varphi^\prime}^2}{ \mu^{\prime 2} }}}\right),
\end{align}
while for $m_{Z^\prime}, m_{\varphi^\prime} \gg \mu^\prime$, it is
\begin{align} \label{eq:ESSlimit_2}
m_3 \simeq& \frac{g^{\prime 2}}{8\pi^2} \frac{m_D^2}{\Lambda^2+m_D^2} \frac{\Lambda^2}{\mu^\prime}  \left( {3 \ln{\frac{m_{Z^\prime}^2}{\Lambda^2}} + \ln{\frac{m_{\varphi^\prime}^2}{\Lambda^2}} - 4 }\right).
\end{align}
In this case, the light neutrino masses are controlled mainly by $\nu_5$, and the intermediate state $\nu_4$ can be much lighter. 
\begin{figure*}[t]
    \centering
    \includegraphics[width=0.49\textwidth]{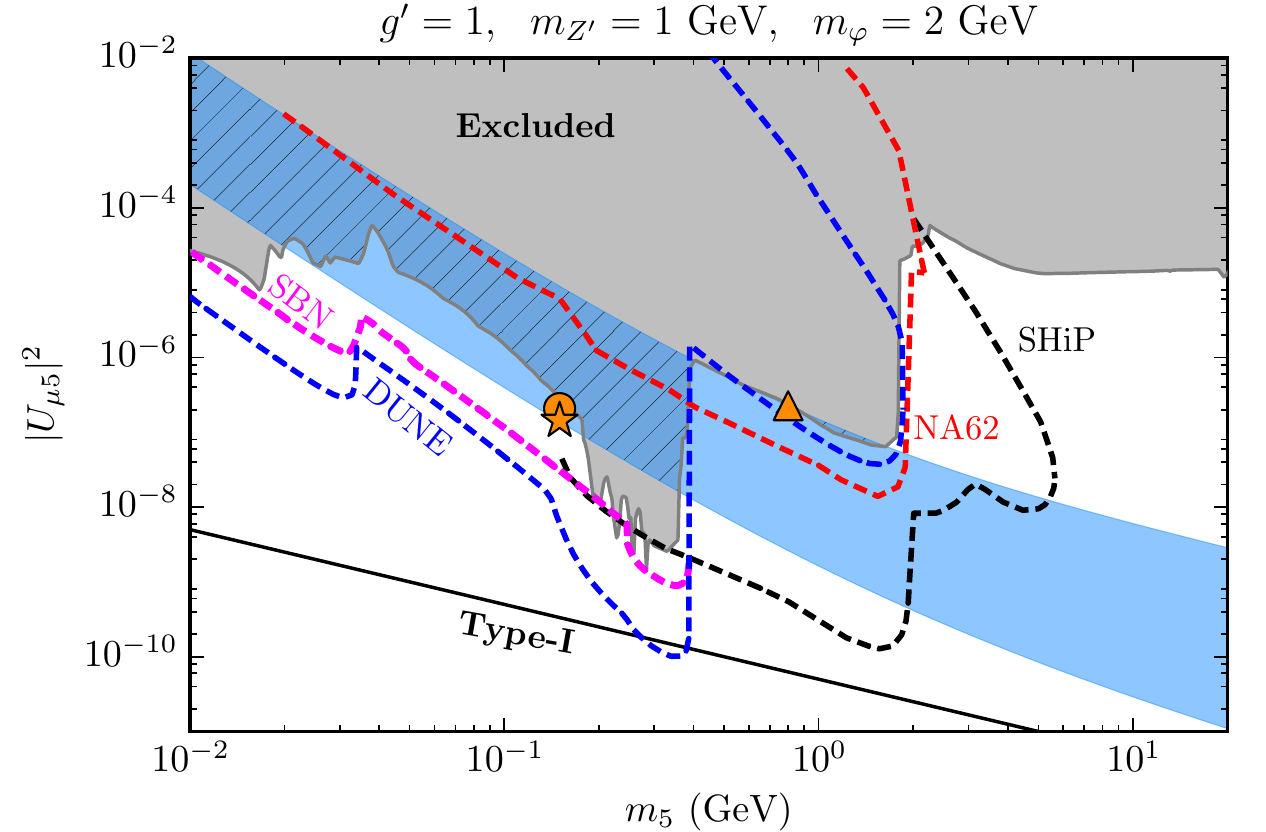}
    \includegraphics[width=0.49\textwidth]{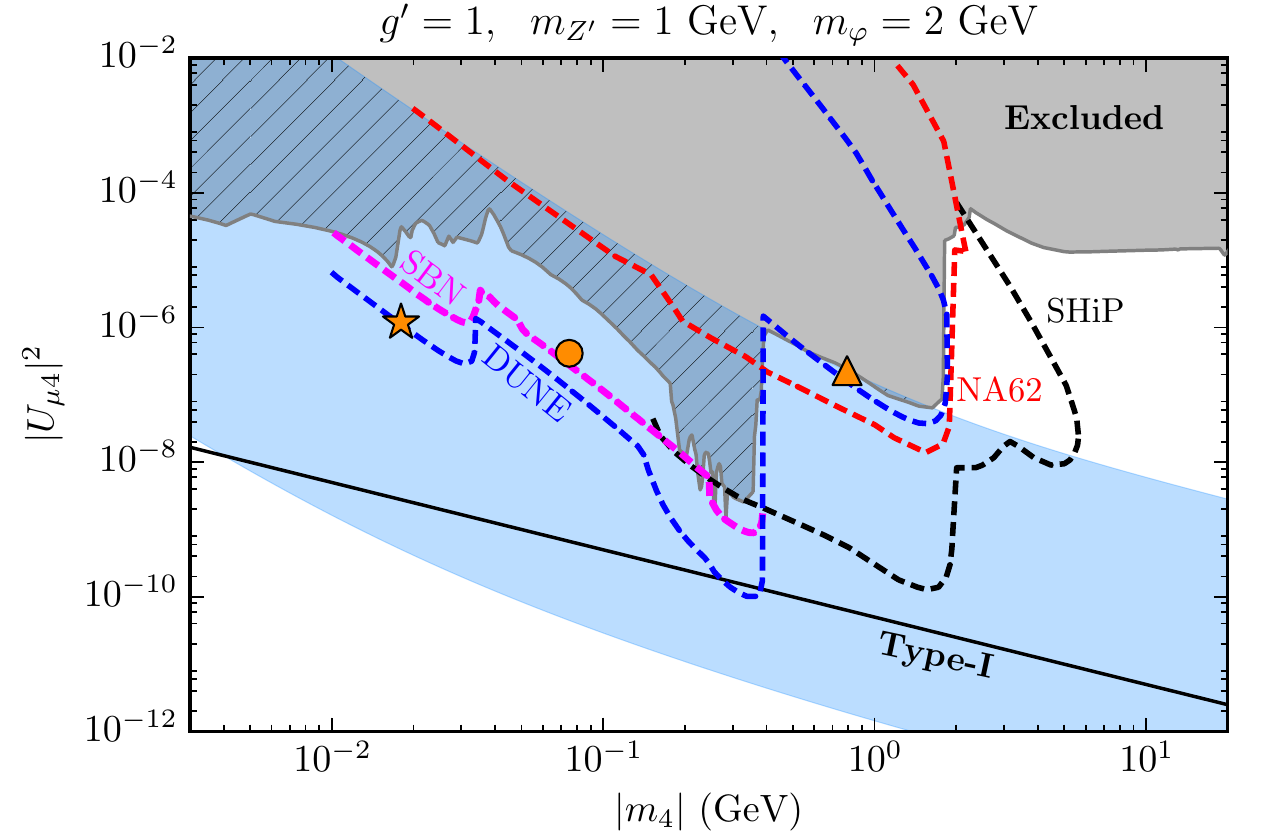}
    \caption{The region of interest for neutrino mass generation in our model in the parameter space of the $\nu_5$ (left) and $\nu_4$ (right) mass states. We require $m_3 = \sqrt{\Delta m^2_{\rm atm}}$ and vary $1\%<m_4/m_5<99\%$. Our BPs are $\bigtriangleup$) $m_5 = 800$ MeV, $m_4/m_5 =99\%$, ${\circ}$) $m_5 = 150$ MeV, $m_4/m_5 =50\%$ and $\star$) $m_5 = 150$ MeV, $m_4/m_5 =12\%$. All bounds and projections displayed assume $\chi=\lambda_{\Phi H}=0$. The dashed black line shows the equivalent Type-I seesaw contribution to the light neutrino mass.\label{fig:mass_constraints}}
\end{figure*}

%%%%%%%%%%%%%%%%%%%%%%%%%%%%%%%%%%%%%%%%%%%%%%%%%%%%%%%%%%%%%%%%%%%%%%%
\section{Searching for the origin of neutrino masses} \label{sec:pure_nu_mixing}

In what follows, we discuss the experimental reach to the heavy neutrinos responsible for neutrino mass generation in our model. Since the vector and scalar portals do not contribute significantly to neutrino masses, we first restrict the study to the case $\chi=\lambda_{\Phi H}=0$. For the sake of simplicity and concreteness, we work with a single generation of light neutrinos and focus on the mixing with the muon neutrino. We emphasise that our model predicts
\begin{equation}
    \frac{m_4}{m_5} = - \frac{U_{\alpha 5}^2}{U_{\alpha 4}^2},
\end{equation}
implying that both heavy neutrinos should be searched for.
For a real mixing matrix one can write $\sum_i^3 U_{D i}^{2} \sim U_{\mu 4}^{2}$ and $U_{D 5}^{2} \sim 1$ for small $U_{\mu 4}$. Using these relations and \refeq{eq:masses_general}, we plot the region of interest for neutrino mass generation in \reffig{fig:mass_constraints}. We require $m_3 = 
\sqrt{\Delta m^2_{\text{atm}}} \sim 0.05~\text{eV}$ and vary $m_4/m_5$ from $1\%$ (ESS-like) to $99\%$ (ISS-like). For the hidden sector parameters, we fix $m_{Z^\prime}=1$~GeV, $m_{\varphi^\prime} = 2$ GeV and $g^\prime = 1$. By decreasing (increasing) the mass of the $Z^\prime$, it is possible to shift the band to smaller (larger) values of the mixing angles, although for values smaller than a few hundred MeV, the neutrino masses have a very mild dependence on $m_{Z^\prime}$ (Eqs.~\ref{eq:ISSlimit_2} and~\ref{eq:ESSlimit_2}). Increasing $m_4/m_5$ to values closer to $100\%$ (\ie\,, decreasing $\mu^\prime$ below $m_5/100$) shifts the top of the band to larger values of mixing angle and asymptotically recovers lepton number as a symmetry. Although this possibility appears excluded for $m_{Z^\prime} = 1$ GeV, it can be achieved by lowering the mass of the mediator particles. For instance, for $m_{Z^\prime} = m_{\varphi^\prime}/2 = 100$ MeV and $m_5 < 100$ MeV, we find that values as small as $\mu^\prime \gtrsim 10^{-3} m_5$ are not covered by the grey region in \reffig{fig:mass_constraints}.
Values of $m_4/m_5 < 1\%$ have no effect in the parameter space of $\nu_5$, since in that limit the $\nu_5$ state (mostly in the $N$ direction) dominates the loop contribution.

The region labelled as excluded in \reffig{fig:mass_constraints} is composed of bounds from peak searches~\cite{
%KEK
Yamazaki:1984sj,
%NA48/2
Artamonov:2014urb,
%NA62
CERNNA48/2:2016tdo}, beam dump \cite{
%PS-191
Bernardi:1985ny,
%CHARM
Bergsma:1983rt,
%NA3
Badier:1986xz,
%NuTeV
Vaitaitis:1999wq, 
%BEBC
CooperSarkar:1985nh, 
%NOMAD
Astier:2001ck} 
and collider experiments~\cite{
Abreu:1996pa,%DELPHI
Akrawy:1990zq,%OPAL
Sirunyan:2018mtv% CMS 2018
}. Current and future neutrino experiments can also cover a large region of parameter space with $m_h \lesssim 2$ GeV. For instance, we show the sensitivity of the Short-Baseline Neutrino program (SBN)~\cite{Ballett:2016opr} and of the Deep Underground Neutrino Experiment (DUNE) near detector~\cite{Ballett:2018fah,Ballett:2019bgd} to heavy neutrinos in decay-in-flight searches. We also show the reach of the NA62 Kaon factory operating in beam dump mode~\cite{Drewes:2018irr}, and the dedicated beam dump experiment Search for Hidden Particles (SHiP)~\cite{Bonivento:2013jag,Alekhin:2015byh}, which will cover a much larger region of parameter space from $400$ MeV to $\lesssim6$ GeV. All bounds and sensitivities shown do not take into account the new invisible decays of the heavy neutrinos. Searches that rely on the visible decay products of the heavy neutrinos need to be revisited if the $\nu_h$ can decay invisibly or if new channels mediated by the vector (and/or scalar) portal dominate. In particular, faster decays of $\nu_h$ can shift decay-in-flight bounds to lower values of mixing angles, as discussed in detail in Ref.~\cite{Ballett:2016opr}. Peak searches apply as shown provided $\nu_h$ does not decay immediately via neutral-current channels with visible charged particles.

Let us first consider the case of subdominant vector and scalar portals. Compared to the ``standard" sterile neutrino case, in which $\nu_h$ have only SM interactions suppressed by neutrino mixing, the new neutral-current interaction can enhance the $\nu_h$ decays into light and heavy neutrinos. A comprehensive analysis is beyond the scope of this article and we focus on three benchmark points (BP) shown in \reffig{fig:mass_constraints} to exemplify the most characteristic properties. The BP represented as a triangle ($\bigtriangleup$) corresponds to $m_5 = 800$ MeV and $m_4/m_5 = 99\%$. In this case, the two heavy states are very degenerate in mass and decay like a ``standard" sterile neutrino via $|U_{\mu 4}|^2$-suppressed SM charge- and neutral-current interactions. The channel $\nu_5 \to \nu_4 \nu_\alpha \overline{\nu}_\alpha$ via the $Z^\prime$ is phase space suppressed and becomes relevant only for larger mass splittings. The invisible $\nu_4$ decay mediated by the $Z^\prime$ is subdominant as it scales as $|U_{\mu 4}|^6$ and becomes important only for larger values of the mixing angles.

For the next BPs we fix $m_5 = 150$ MeV. If we take $m_4/m_5 = 50\%$, as we do for the BP represented by the circle ($\circ$), $\nu_5$ will predominantly decay to $\nu_4 \nu_\alpha \overline{\nu}_\alpha$ due to the $Z^\prime$ contribution (provided $|U_{\mu 5}|^2\gtrsim \left( m_{Z^\prime}/m_{Z} \right)^4$). Consequently, the best candidate for detection is the $\nu_4$ via the SM weak decays $\nu_4 \to \nu_\alpha e^+e^-$. The values of the mixing angles for this BP, $|U_{\mu 4}|^2 \sim 3\times 10^{-7}$ and $|U_{\mu 5}|^2 \sim 10^{-7}$, are within reach of the SBN and DUNE experiments. 
For a larger mass hierarchy, e.g. $m_4/m_5 = 12\%$, see star BP ($\star$), the $Z^\prime$ mediated decay $\nu_5 \to \nu_4 \overline{\nu_4} \nu_4$ dominates, inducing a large $\nu_4$ population in addition to the states already produced in the beam. The intermediate state $\nu_4$ can further decay as in the previous case into $\nu_4 \to \nu_\alpha e^+e^-$. For the mixing angles we are considering, $|U_{\mu 4}|^2 \sim 10^{-6}$ and $|U_{\mu 5}|^2 \sim 10^{-7}$, DUNE will be able to test this BP. Similar considerations apply to the case where $m_5 > m_4 + m_{Z^\prime}$, where now the $Z^\prime$ can be produced on-shell in the $\nu_5$ decay. The behaviour of $\nu_4$ is as discussed above. If $ m_{Z^\prime} < m_4$, then both heavy neutrinos predominantly decay into neutrinos and the $Z^\prime$, which presents a challenge for detection as it produces mainly light neutrinos.

Experimental detection of the $Z^\prime$ and $\varphi^\prime$ particles in the absence of kinetic and scalar mixing is also daunting. Nevertheless, they can be searched for in the kinematics of charged particles from meson decays~\cite{Laha:2013xua,Bakhti:2017jhm}. Another strategy is to search for the neutrino byproducts of the decay of a $Z^\prime$ produced at accelerator neutrino facilities~\cite{Bakhti:2018avv}. 

If the vector (and scalar) portals are non-negligible, the phenomenology could be significantly richer, as discussed in \cite{Ballett:2019pyw}. In particular, $Z^\prime$-mediated decays into $\nu_\alpha e^+ e^-$, and $\nu_\alpha \mu^+ \mu^-$ if kinematically allowed, could dominate even for tiny values of $\chi^2$. For instance, for the circle BP, $\chi^2 $ as low as $10^{-8}$ would make the above decays the main channels. Pseudo-scalar final states are suppressed due to the vector nature of the $Z^\prime$. The scalar portal is expected to give subdominant contributions due to the small Higgs-electron Yukawa coupling, although decay chains with intermediate $\nu_4$ states may become relevant. Finally, cosmological bounds on heavy neutrino in the 10 MeV -- GeV scale may be weakened as they would decay well before Big Bang Nucleosynthesis~\cite{Dolgov:2000jw} (see also the discussion in Ref.~\cite{Hannestad:2013ana,*Dasgupta:2013zpn,*Mirizzi:2014ama,*Chu:2015ipa,*Cherry:2016jol,*Chu:2018gxk,*Song:2018zyl})%The BR would have a very different structure compared to the standard neutrino and vector portals. By looking for the decay channels, one would be able to, at least partially, disentangle the neutrino, vector and scalar contributions.

We have focused on the mixing with muon neutrinos as these provide one of the most sensitive avenue to test the model. In the electron sector, direct bounds on the active-heavy mixing are similar, with peak searches from $\pi^\pm$ decay being most relevant below $\approx 100$ MeV. For cases with large LNV, heavy neutrinos can dominate neutrinoless double beta decay~\cite{LopezPavon:2012zg}, and this sets the strongest constraints in the parameter space. The tau sector is relatively poorly constrained, so greater freedom exists if such entries are relevant for neutrino mass generation. 

\section{Conclusions} We have considered a recently proposed model which invokes the existence of a dark neutrino sector and a new hidden gauge symmetry, focusing on the generation of neutrino masses in this context.
The presence of a hidden broken gauge symmetry protects the neutrino mass matrix, leading to a Minimal ISS-like structure, and allows for one-loop diagrams involving the new vector and scalar content to generate the correct neutrino masses. Searches for the neutrino mass generation in our model are possible via conventional heavy neutral lepton searches, as well as through exotic signatures arising from the interplay of portal couplings.

\section*{Acknowledgements}
The authors would like to thank Kris Moffatt for insights into the neutrino mass generation mechanism. This work was partially supported by Conselho Nacional de Ci\^{e}ncia e Tecnologia (CNPq). This project also received support from the European Union's Horizon 2020 research and innovation programme under the Marie Sk\l odowska-Curie grant agreement No. 690575 (RISE InvisiblesPlus) and No. 674896 (ITN Elusives). SP and PB are supported by the European Research Council under ERC Grant NuMass (FP7-IDEAS-ERC ERC-CG 617143). SP acknowledges partial support from the Wolfson Foundation and the Royal Society.

%%%%%%%%%%%%%%%%%%%%%%%%%%%%%%%%%%%%%%%%%
\bibliographystyle{apsrev4-1}
\bibliography{main}{}
%%%%%%%%%%%%%%%%%%%%%%%%%%%%%%%%%%%%%%%%%

\end{document}